\documentclass[twocolumn,showpacs,preprintnumbers,amsmath,amssymb]{revtex4-1}
\usepackage{graphicx}
\usepackage{dcolumn}
\usepackage{bm}
\usepackage{color}
\begin{document}
\title{Hexagonal boron nitride is an indirect bandgap semiconductor}
\author{G. Cassabois, P. Valvin, B. Gil$^{\ast}$}
\affiliation{Laboratoire Charles Coulomb (L2C), UMR 5221 CNRS-Universit\'e de Montpellier, F-34095, Montpellier, France}
\date{\today}
\begin{abstract}
Hexagonal boron nitride is a wide bandgap semiconductor with a very high thermal and chemical stability often used in devices operating under extreme conditions. The growth of high-purity crystals has recently revealed the potential of this material for deep ultraviolet emission, with an intense emission around 215 nm. In the last few years, hexagonal boron nitride has been raising even more attention with the emergence of two-dimensional atomic crystals and Van der Waals heterostructures, initiated with the discovery of graphene. Despite this growing interest and a seemingly simple structure, the basic questions of the bandgap nature and value are still controversial. Here, we resolve this long-debated issue by bringing the evidence for an indirect bandgap at 5.955 eV by means of optical spectroscopy. We demonstrate the existence of phonon-assisted optical transitions, and we measure an exciton binding energy of about 130 meV by two-photon spectroscopy.
\end{abstract}
\maketitle
Hexagonal boron nitride (hBN) exhibits unique electronic properties such as a wide bandgap, low dielectric constant, high thermal conductivity, and chemical inertness. In contrast to other nitride semiconductors such as GaN and AlN \cite{nanishi} which most stable crystalline phase is of the wurtzite type, the hexagonal structure of hBN makes it a prototype 2D material along with graphene and molybdene disulfide \cite{karnik}. With a honeycomb structure based on sp$^2$ covalent bonds similar to graphene, bulk hBN has first gained tremendous attention as an exceptional substrate for graphene with an atomically smooth surface. 2D hBN or 'white graphene', in the form of few-layer crystals or monolayers of hBN, has then appeared as a fundamental building block of Van der Waals heterostructures \cite{geim}.

In spite of this rising interest for hBN and the large number of studies devoted to this material of seemingly simple crystal structure, the very basic question of the bandgap nature is still controversial. There is a strong contrast between ab initio band structure calculations predicting an indirect bandgap crystal \cite{xu,furthmuller,blase,arnaud,gao} and optical measurements concluding to a direct one \cite{watanabe04,zunger,evans}. In 2004, Watanabe \textit{et al.} showed in particular that hBN is a very promising material for light-emitting devices in the deep ultraviolet domain, with an intense luminescence peak at 5.76 eV, supporting the direct nature of the bandgap \cite{watanabe04}. The use of high-purity hBN crystals has then allowed the demonstration of lasing at 215 nm by accelerated electron excitation \cite{watanabe04}, and also the operation of field emitter display-type devices in the deep ultraviolet \cite{watanabe09b,watanabe11}.

Here, we demonstrate that hBN has an indirect bandgap at 5.955 eV and that the optical properties of hBN are profoundly determined by phonon-assisted transitions. By means of two-photon spectroscopy, we reveal the existence of previously unobserved lines. The weakest one lies at the highest energy (5.955 eV) and it corresponds to the dim emission of the indirect exciton. Each emission line appearing at lower energy consists in a phonon replica, for which we identify the corresponding phonon mode. Finally, two-photon excitation spectroscopy allows us to measure, for the first time in an indirect bandgap semiconductor, the energy splitting between the $1s$ and $2p$ exciton states. We obtain an estimation of the exciton binding energy of 128$\pm$15 meV, showing that excitons in hBN are of Wannier type, and that the single-particle bandgap is at an energy of 6.08$\pm$0.015 eV in hBN.

\begin{center}
\begin{figure}[h,t]
\centering
  \begin{tabular}{@{}cc@{}}
  \multicolumn{2}{c}{\includegraphics[width=.48\textwidth]{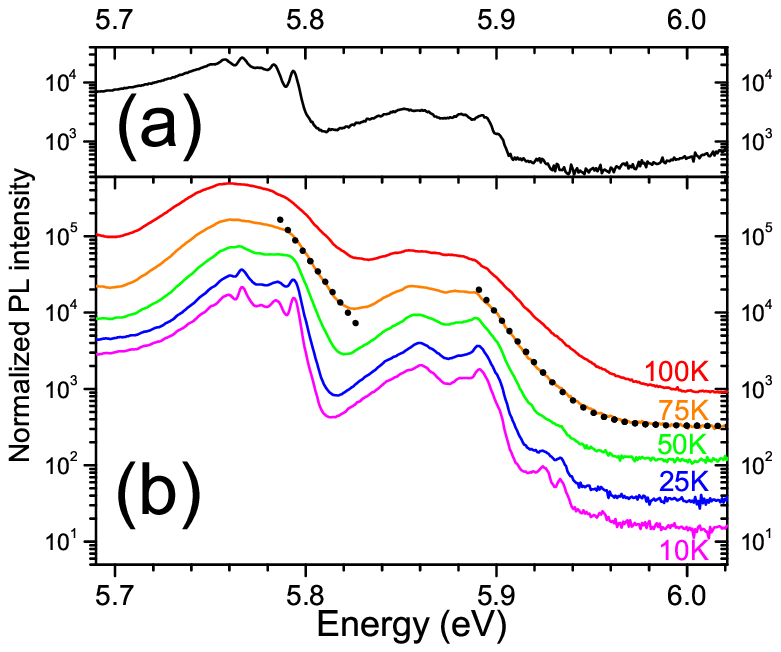}}\\
    \includegraphics[width=.22\textwidth]{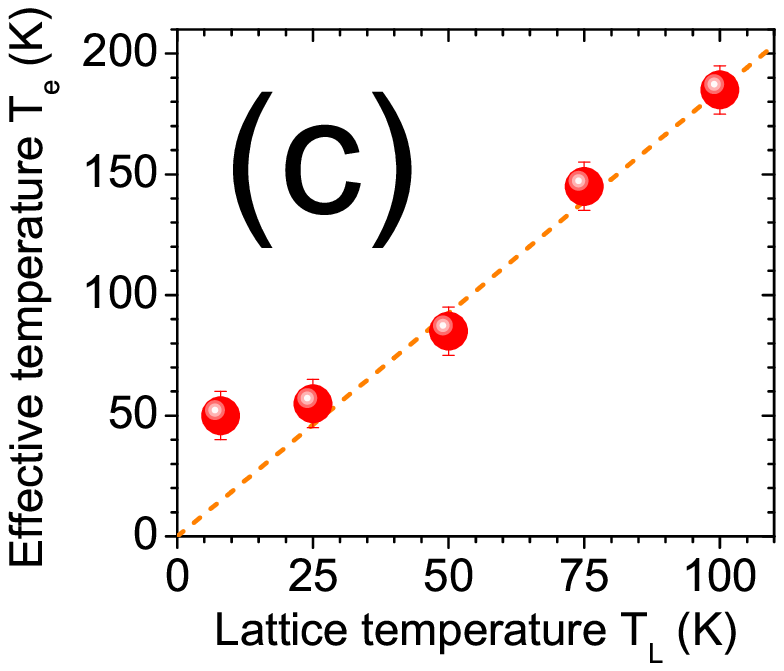} &
    \includegraphics[width=.24\textwidth]{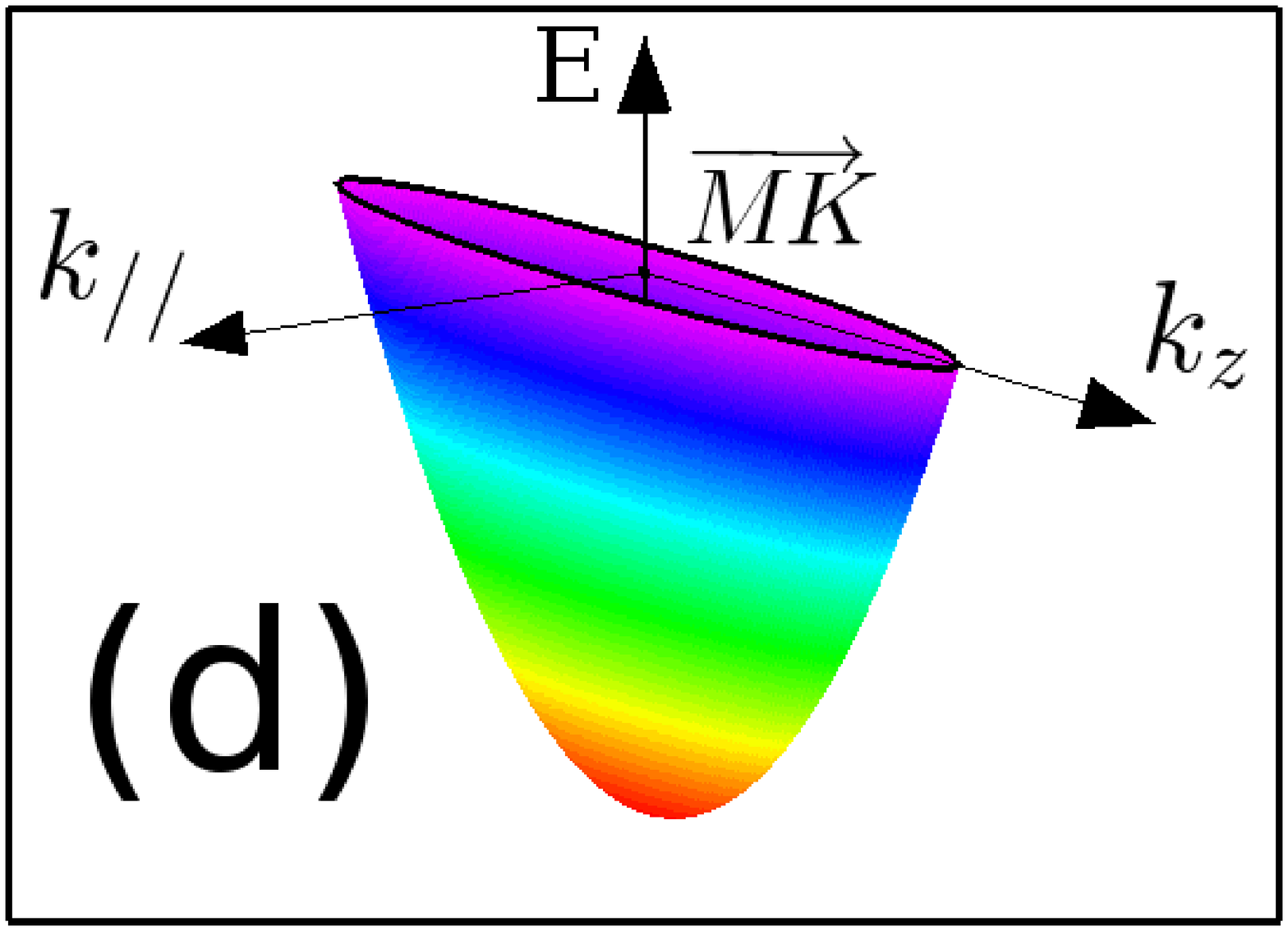}
  \end{tabular}
\caption{\textbf{Monitoring exciton thermalization in phonon replicas}. (a) Photoluminescence (PL) spectrum of hexagonal boron nitride at 10K for a one-photon excitation at 6.3 eV. (b) PL spectrum of hexagonal boron nitride, for a two-photon excitation at 3.03 eV, as a function of temperature (the spectra are shifted for clarity). The dotted lines indicate a Boltzmann law with an effective temperature $T_e$=145K superimposed on a constant baseline. (c) Effective temperature $T_e$ (symbols) as a function of lattice temperature $T_L$. The dashed curve indicates a linear regression with a 1.85 slope. (d) Schematic representation of the anisotropic energy dispersion of the indirect exciton $iX$ around the $\protect\overrightarrow{MK}$ point of the reciprocal space. The colormap corresponds to a thermal distribution of indirect excitons, with a density decreasing from red to blue.}
\label{fig1}
\end{figure}
\end{center}

In Fig.~\ref{fig1}(a) we display a photoluminescence (PL) spectrum of hBN at low temperature, in the usual configuration of one-photon excitation at 6.3 eV. We observe the luminescence peak at 5.76 eV reported by Watanabe \textit{et al.} in high-purity samples and attributed to the recombination of free excitons \cite{watanabe04}. Spatially-resolved cathodoluminescence measurements have confirmed this interpretation on the basis of the homogeneous spatial distribution of the emission intensity in hBN crystallites \cite{jaffrenou07,watanabe11b} and in few-layer hBN flakes \cite{pierret,bourrelier}. The presence of defects in hBN leads to two additional emission bands centered at 5.5 eV and 4 eV \cite{silly,jaffrenou07,museur08,museur08b,watanabe11}. In contrast to the 5.76 eV emission line of the free exciton, these defect-related emission bands display strong localization near dislocations and boundaries in cathodoluminescence measurements \cite{jaffrenou07,watanabe11b,pierret,bourrelier}, with a striking spatial anti-correlation with the free exciton PL at 5.76 eV, as recently characterized with nanometric resolution in a transmission electron microscope \cite{bourrelier}.

In Fig.~\ref{fig1}(a), one also observes that the 5.76 eV emission line is a multiplet with fine structures extending over 40 meV, accompanied by a similar satellite band at 5.86 eV of lower intensity \cite{watanabe09,watanabe11b,pierret}. This satellite band shows the same delocalized emission over the hBN crystals as the 5.76 eV one, characteristic of free exciton recombination \cite{bourrelier}. The different free-exciton levels observed in Fig.~\ref{fig1}(a) (and usually called $S$ lines in the literature) have been tentatively attributed to dark and bright excitons with a degeneracy lifted by a Jahn-Teller effect \cite{watanabe09}. Still, all theoretical calculations predict an indirect bandgap for hBN \cite{xu,furthmuller,blase,arnaud,gao}. Moreover, our two-photon excitation scheme allows us to detect two previously unreported lines, namely a weak doublet around 5.93 eV, and an even dimmer line at 5.955 eV (Fig.~\ref{fig1}(b)), thanks to the suppression of the background due to laser light scattering in one-photon spectroscopy (Fig.~\ref{fig1}(a)). In the following, we will provide a comprehensive understanding of the opto-electronic properties of hBN in the deep ultraviolet, which display all the features of an indirect bandgap semiconductor.

Ab initio calculations predict an indirect bandgap for hBN with extrema of the band structure located around the M and K points of the Brillouin zone for the conduction and valence bands, respectively \cite{xu,furthmuller,blase,arnaud,gao}. Excitonic effects modify the single-particle picture of band structure calculations, whatever the direct or indirect nature of the optical transition \cite{bassani}. As in other indirect semiconductors \cite{handbook}, the hBN indirect exciton ($iX$) corresponding to the electron-hole pair built around the M and K points of the Brillouin zone is thus not coupled to light in the dipolar approximation, and phonon scattering is required in order to fulfill momentum conservation during photon emission or absorption \cite{cho}.

The first evidence for recombination assisted by phonon emission in hBN arises from the observation of a thermal distribution of excitons in the high-energy tail of the different emission lines. In Fig.~\ref{fig1}(b), we display the PL spectrum of hBN as a function of temperature, for a two-photon excitation at 3.03 eV. Thanks to our background-free excitation scheme, we see that, for both the 5.76 and 5.86 eV emission lines, the PL signal intensity falls exponentially on their high-energy side, with a slope decreasing on raising the temperature. On the contrary, on their low-energy side, these emission bands remain unchanged with temperature, except for some residual contamination by the intense red-shifted neighbouring lines. In order to quantitatively analyze the exponential decrease of the PL signal at high energy, we have fitted our data with Boltzmann distributions of effective temperature $T_e$. As shown for instance at 75K in Fig.~\ref{fig1}(b), we obtain an excellent agreement, with systematically the same effective temperature for the 5.76 and 5.86 eV emission lines. The effective temperature $T_e$ is further plotted as a function of the lattice temperature $T_L$ in Fig.~\ref{fig1}(c), and for temperatures larger than 25K, we observe a thermalization of the excitonic system with the surrounding crystal.

Such a phenomenology is typical of semiconductor materials. Hot excitons are initially created with a large kinetic energy after electrical or optical excitation. During carrier relaxation their energy distribution converges to a thermal Boltzmann law via phonon-assisted scattering processes. However, the exciton thermalization can not be observed in the so-called zero-phonon line since only excitons with wavevector close to zero can contribute to the PL signal by direct emission of photon. The population of free excitons with large wavevectors can only be monitored by studying phonon replicas because phonon emission ensures momentum conservation in their radiative recombination. This universal effect was observed in many different semiconductors \cite{weisbuch,snoke,umlauff,xu06}. In the specific context of exciton condensation, the transition from a Boltzmann to a Bose-Einstein distribution is indeed investigated by a careful study of the high-energy tail of phonon replicas \cite{snoke2}. Since only phonon-assisted processes can give an accurate replica of the energy distribution of excitons, we first conclude that the observation of a thermal distribution of excitons in hBN is a first piece of evidence for the nature of the 5.76 and 5.86 eV emission lines as being phonon replicas.

As far as the thermalization process is concerned, the strong structural anisotropy of hBN translates in an effective temperature $T_e$ of the Boltzamnn law larger than the lattice temperature $T_L$, as can be seen in Fig.~\ref{fig1}(c) where the linear regression (dashed curve) has a slope of 1.85. Although striking at first sight, this effect simply arises when taking into account the k-dependence of the phonon energy in the phonon-assisted recombination process (Supplementary information). While being usually negligible in other semiconductors leading to $T_e\sim T_L$ \cite{snoke2}, this correction is important in hBN because of the flatness of both phonon and electron dispersions along the z-axis \cite{arnaud,serrano}. In fact, the corresponding large effective mass (as schematically shown in Fig.~\ref{fig1}(d)) leads to a huge increase of the density of exciton states, so that the phonon replicas mostly monitor the exciton thermalization along the z-axis. Such a situation is in strong analogy with the case of quantum well superlattices, where the two-dimensional excitons thermalize and efficiently redistribute over the whole superlattice mini-zone \cite{shtrichman}.
\begin{center}
\begin{figure}[h,t]
\centering
  \begin{tabular}{@{}cc@{}}
  \includegraphics[width=.26\textwidth]{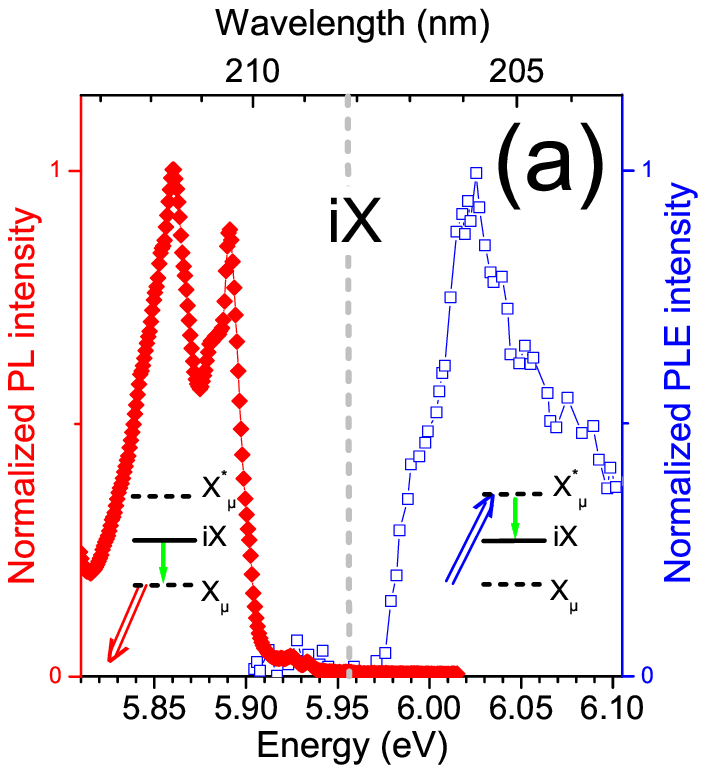} &
    \includegraphics[width=.215\textwidth]{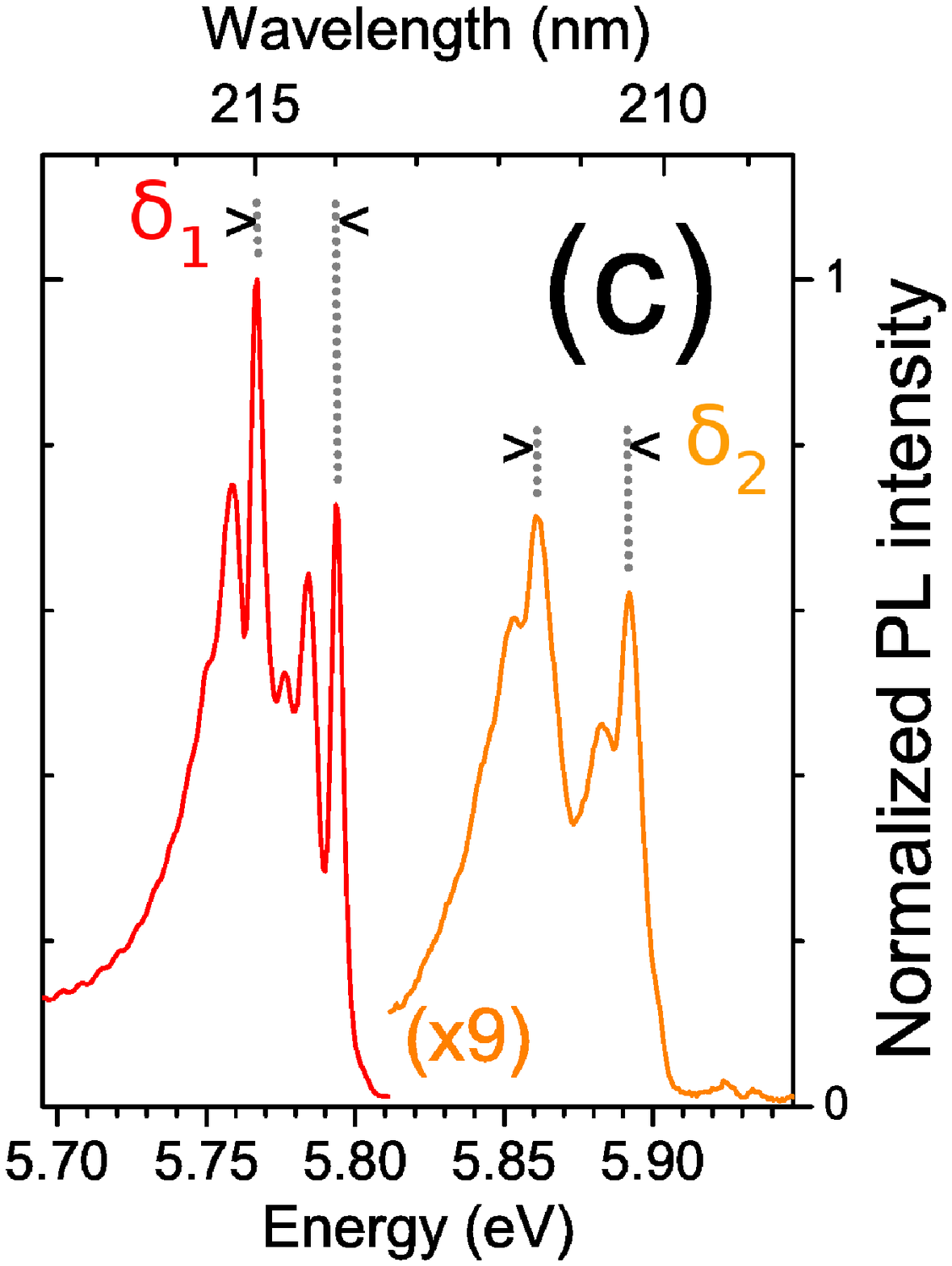}\\
   \multicolumn{2}{c}{\includegraphics[width=.46\textwidth]{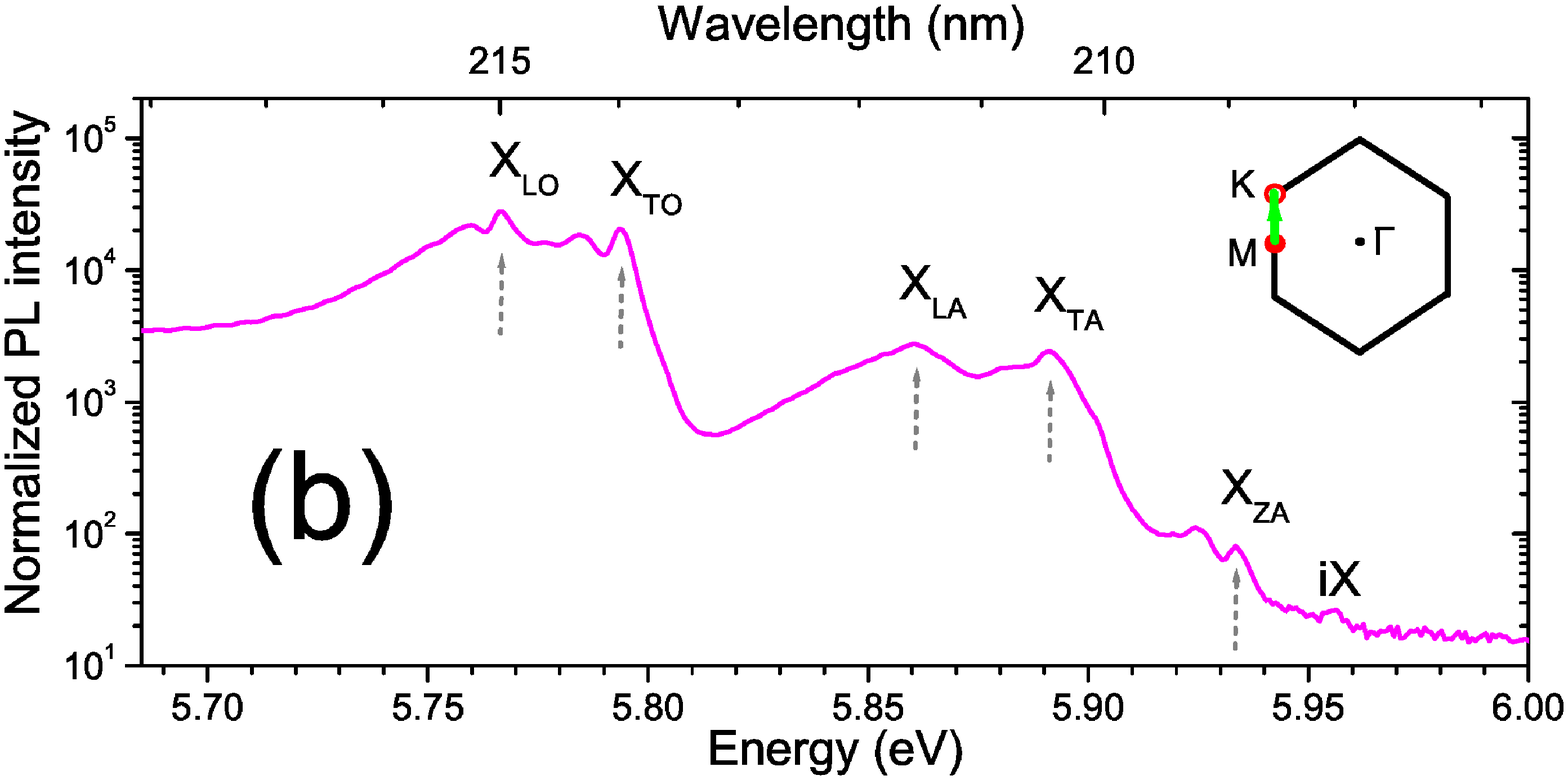}}
  \end{tabular}
\caption{\textbf{Phonon-assisted emission and absorption in hexagonal boron nitride}. (a) Normalized photoluminescence (PL) signal intensity (red diamonds) at 10K as a function of detection energy, and normalized photoluminescence excitation (PLE) signal intensity (blue open squares) at 11K (from Ref.~\cite{museur11}) as a function of excitation energy. Schematic representations of recombination (red double-arrow) and absorption (blue double-arrow) assisted by phonon emission (vertical green arrow) involving the virtual states $X_{\mu}$ and $X^*_{\mu}$, respectively, where $\mu$ is a phonon mode. (b) Identification of the phonon modes involved in the phonon-assisted recombination lines in hBN. The scattering path in the first Brillouin zone is indicated by the green arrow, corresponding to the phonons wavevector. (c) Normalized PL spectrum at 10K for a two-photon excitation at 3.03 eV, at low excitation power for optimal observation of the fine structures.}
\label{fig2}
\end{figure}
\end{center}

\begin{center}
\begin{figure}[h,t]
    \includegraphics[scale=0.8]{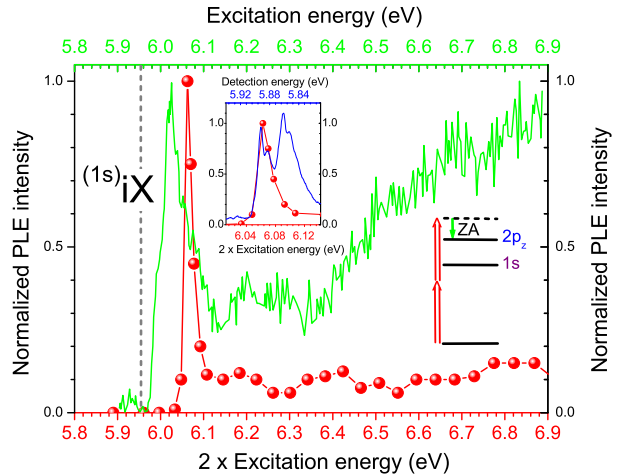}
\caption{\textbf{Two-photon excitation spectroscopy in hexagonal boron nitride}. Normalized photoluminescence signal intensity at 5.76 eV as a function of twice the excitation energy (red circles) for two-photon spectroscopy, and as a function of the excitation energy (solid green line) for one-photon spectroscopy (from Ref.~\cite{museur11}), at 10K. Inset: superposition of the two-photon excitation spectrum and PL spectrum, for a same 115 meV-range, but for increasing and decreasing energy, respectively: in contrast to phonon-assisted recombination, only one phonon mode is involved in the two-photon excitation spectrum. Scheme of the two-photon excitation of the $2p_z$ exciton state assisted by the emission of a ZA phonon.}
\label{fig3}
\end{figure}
\end{center}

The 5.76 and 5.86 eV emission lines being identified as phonon replicas, we now turn to the determination of the indirect exciton $iX$ energy. In the optical response of indirect semiconductors, there is a mirror symmetry between the two processes of absorption and emission assisted by phonon emission \cite{cho}: compared to the bandgap energy, photon emission is redshifted by the phonon energy while photon absorption is blue-shifted by the same value (see energy level schemes in Fig.~\ref{fig2}(a)). We will show now that the emission line of smallest intensity, observed at 5.955 eV in Fig.~\ref{fig1}(b), corresponds to the indirect exciton $iX$. In order to check the mirror image between emission and absorption around 5.955 eV, we have plotted in Fig.~\ref{fig2}(a) the normalized PL spectrum (red diamonds) and the normalized PLE spectrum (blue open squares) in a spectral window centered at 5.955 eV. It is essential to confront the PL spectrum of free excitons to the PLE one detected at their corresponding emission energy, as it is the case in Ref.\cite{museur11}, but not in previous studies reporting PLE spectroscopy for detection windows centered at defect-related lines \cite{evans,jaffrenou08,museur08,museur08b}. In Fig.~\ref{fig2}(a), we observe that the signal intensity is of the order of the noise around 5.955 eV in both cases, and that it increases to its maximum value by a blue-shift (red-shift) of 60 meV in the PLE (PL, respectively) spectrum. Moreover, the lines have a half-width at half-maximum (HWHM) of the order of 25 meV in both emission and absorption spectra. Although the rise of the PLE signal around 6.01 eV is not as steep as in the PL spectrum around 5.9 eV (which is due to a poorer spectral resolution in the one-photon excitation spectroscopy in Ref.~\cite{museur11}), we will show later by two-photon excitation spectroscopy (Fig.~\ref{fig3}, inset), that the PLE signal has the same abrupt rise as the PL signal, thus confirming the mirror image of emission and absorption around 5.955 eV.

In the simplest case of one-phonon scattering processes, momentum conservation completely determines the phonon wavevector as \textbf{MK} (corresponding to the green arrow in the first Brillouin zone in Fig.~\ref{fig2}(b)). From simple geometrical considerations of the hexagonal Brillouin zone, one deduces that one-phonon scattering processes involve phonons in the middle of the Brillouin zone, around $\Sigma$ points \cite{reich}. Consequently, the energy detuning between the phonon replicas and the indirect exciton $iX$ is to reflect the energy of the different phonon modes around $\Sigma$ points. In Fig.~\ref{fig2}(b), we have indicated by vertical arrows the position of the five doublet lines observed in the PL spectrum. Their energy detunings with $iX$ are 22, 64, 95, 162 and 188$\pm$1 meV, respectively. From the complete phonon band structure characterization reported in Ref.\cite{serrano}, we observe a perfect agreement with the phonon energy in the middle of the Brillouin zone for the ZA, TA, LA, TO and LO modes, respectively. The corresponding assignment displayed in Fig.~\ref{fig2}(b) thus allows us to identify the intense emission band at 5.76 eV as due to two optical phonon replicas, whereas the less intense emission between 5.8 and 5.94 eV arises from three acoustic phonon replicas. This situation corresponds to the general phenomenology observed in semiconductor materials where the electron-phonon coupling is much more efficient in the case of optical phonons for which the coupling is given by the Fr\"{o}hlich interaction, while the deformation potential and piezoelectric coupling lead to less efficient scattering processes in the case of acoustic phonons \cite{cho}. As far as the acoustic phonon replicas are concerned, we explain the low signal intensity of the $X_{ZA}$ line as a result of the selection rule controlling radiative recombination assisted by phonon in hBN, which is forbidden by symmetry for a ZA phonon mode in hBN in our experimental configuration (Supplementary information).

In the light of this identification of five phonon replicas in the PL spectrum of hBN, we also interpret the varying visibility of the doublet structure in each replica as being due to the group velocity of the corresponding phonon branch. For acoustic phonons, the group velocity increases when passing from ZA to TA and LA modes \cite{serrano} so that momentum conservation fulfilled within a given k-space interval results in an increasing energy variation, thus smoothing the fine structure of the phonon replicas. This effect is more clearly observed in Fig.~\ref{fig2}(c) for which the PL spectrum is plotted on a linear scale, and where the contrast of the doublet structure decreases from the ZA to LA replica. On the contrary, optical phonons have a smaller group velocity than acoustic phonons \cite{serrano} therefore revealing with maximum contrast the fine structure of a given phonon replica (Fig.~\ref{fig2}(c)). Incidentally, we note that a triplet structure develops for optical phonon replicas, and we speculate the existence of a zone-folding effect due to the presence of multi-layer segments of different thickness in our sample, in analogy to the observation performed in multi-layer graphene \cite{lui}. Finally, we highlight that the energy detunings $\delta_1$=31$\pm$1 meV and $\delta_2$=26$\pm$1 meV (Fig.~\ref{fig2}(c)) discussed in the literature \cite{watanabe09} simply reflect the TO-LO energy splitting and the TA-LA one in the middle of the Brillouin zone.

We thus bring the demonstration that hBN does have an indirect bandgap, in agreement with theoretical calculations \cite{xu,furthmuller,blase,arnaud,gao}. Furthermore, our estimation of 5.955 eV is fully consistent with electron energy loss spectroscopy in hBN, where electronic momentum transfer allows the excitation of indirect bandgap material, in contrast to optical spectroscopy. The bandgap of bulk hBN was found to be 5.9$\pm$0.2 eV \cite{tarrio} and in multi-wall boron nitride nanotubes in which quantum confinement is negligible, a value of 5.8$\pm$0.2 eV \cite{arenal} was obtained.

In the prospect of further elucidating the fundamental opto-electronic properties of hBN, we have performed two-photon excitation spectroscopy in order to determine the exciton binding energy. Two-photon spectroscopy is a powerful technique for accessing information on the bound states of the exciton relative motion \cite{bassani}, which has recently regained attention in carbon nanotubes \cite{wang,maultzsch} or transition metal dichalcogenide monolayers \cite{ye,wangtoulouse}. In the case of direct bandgap materials, the optical selection rules for two-photon absorption impose the excitation of p-exciton states, thus providing an estimation of the 1s-2p energy splitting. The case of indirect bandgap compounds is less documented, with the evidence for phonon-assisted two-photon absorption only in germanium \cite{tuncel}, to the best of our knowledge.

In Fig.~\ref{fig3}, we display, in red circles, the two-photon excitation spectrum in hBN for a detection at 5.86 eV as a function of twice the excitation energy. Our data are compared to the one-photon excitation spectrum (solid green line) from Ref.~\cite{museur11}. Below 6.02 eV, the two-photon excitation does not create any carrier in the hBN sample. Starting from 6.04 eV, there is an abrupt rise of the PLE signal intensity, which is maximum at 6.062 eV and then decreases by one order of magnitude to an approximately constant value up to 7 eV. From the comparison of the one-photon and two-photon PLE spectra, we first conclude that there is a 60 meV blue-shift of the PLE maximum in the nonlinear measurements by two-photon spectroscopy. Moreover, we observe that the resonance in the two-photon PLE spectrum is very narrow, with a HWHM of only 10 meV. This is half the value of the 5.76 and 5.86 eV emission lines, that consist in two phonon replicas (Fig.~\ref{fig2}(b)), therefore suggesting the implication of a single phonon mode in two-photon absorption. In fact, in the inset of Fig.~\ref{fig3} where we plot the two-photon PLE spectrum and the PL spectrum, for a same 115 meV-range but for increasing and decreasing energy, respectively, we observe that the resonance in the two-photon PLE spectrum is indeed the mirror image of a single phonon replica with the same abrupt rise followed by a smoother decrease as a function of energy.

We thus conclude that two-photon excitation is assisted by a single phonon mode in hBN. This situation contrasts with the optical response in emission, where four phonon modes (TA, LA, TO and LO) give rise to prominent replicas, the symmetry-forbidden ZA mode leading to a weak PL line. As shown in the Supplementary information, the selection rules for phonon-assisted two-photon absorption indicate on the contrary that the ZA phonon mode is allowed, for excitation of the 2p$_z$ exciton state, while the acoustic phonons of higher energy (TA and LA modes) contribute to the absorption of the 2p$_{x,y}$ exciton states. Since the ZA mode corresponds to the phonon of lowest energy, we tentatively attribute the sharp line in the two-photon PLE spectrum as being assisted by a ZA phonon. Given the estimated value of 22$\pm$1 meV for the ZA phonon energy (Fig.~\ref{fig2}(b)), we obtain an energy splitting of 85$\pm$1 meV between the 1s and 2p$_z$ states of the indirect exciton $iX$ in hBN. This value is much smaller than the bandgap energy and it shows that excitons in hBN are of Wannier type, in contrast to theoretical calculations predicting Frenkel excitons \cite{arnaud}.

Under the assumption of the Rydberg series usually observed for Wannier excitons, such a splitting would lead to an exciton binding energy of 113 meV in an isotropic material. However, because of the strong anisotropy of hBN, one has to correct this value in the framework of the theory developed for anisotropic excitons in semiconductors \cite{baldereschi,gil}. Depending on the value of the so-called anisotropy factor $\gamma$ given by the ratio, for the in- and out-of-plane cases, of the dielectric constant times effective mass, one finds a 1s-2p$_z$ splitting ranging from 0.75 (for $\gamma$=1) to 0.6 (for $\gamma$=0.1) in unit of the excitonic Rydberg \cite{gil}, therefore giving an upper bound of 142 meV for the exciton binding energy in the limit of strong anisotropy ($\gamma$=0.1). The mean value of 128$\pm$15 meV is larger than the exciton binding energy in diamond (70 meV) \cite{handbook} or AlN (52 meV) \cite{gil} but still far from a Frenkel exciton.

Eventually, we estimate the single-particle bandgap of hBN at 6.08$\pm$0.015 eV. This value indeed corresponds to the onset of the large absorption band in the one-photon PLE spectrum (Fig.~\ref{fig3}). As a consequence, the $X^*_{TO}$ and $X^*_{LO}$ virtual states at 6.12 and 6.14 eV, respectively, are resonant with the continuum, thus preventing their observation in the one-photon PLE spectrum. It also explains a posteriori why the two-photon PLE spectrum displays only the ZA phonon mode, since the next expected one, corresponding to the TA mode, lies at 6.1 eV, \textit{i.e.} already above the continuum onset. 

In conclusion, we have resolved the long-debated issue of the bandgap nature of hBN by demonstrating that the bandgap is indirect with a value of 5.955 eV. We have shown that the emission spectrum of hBN in the deep ultraviolet is profoundly structured by phonon-assisted recombination and we have identified the various phonon modes in the replicas. We have performed phonon-assisted two-photon absorption from which we have derived an exciton binding energy of about 130 meV, thus revealing that the indirect excitons in hBN are of Wannier type, and that the single particle bandgap in hBN is about 6.08 eV. We highlight the need for theoretical calculations explaining the efficient exciton-phonon interaction in hBN. We hope our results will stimulate experiments addressing these questions, with the exciting possibility to study either the 3D case in high-purity crystals or the 2D one in hBN monolayers, where a transition to a direct bandgap is further expected as in transition metal dichalcogenide compounds.


\textbf{Methods}

\textit{Sample and experimental setup}

Our sample is a commercial hBN crystal from HQ Graphene (http://www.hqgraphene.com/). In our experimental setup, the sample is hold on the cold finger of a closed-circle cryostat for temperature-dependent measurements from 10K to room temperature. Optical excitation is performed at normal incidence.  In the standard configuration of one-photon excitation, the excitation beam is provided by the fourth harmonic of a cw mode-locked Ti:Sa oscillator with a repetition of 82 MHz, and in the case of two-photon excitation by the second harmonic of the Ti:Sa oscillator. The spot diameter is of the order of 200 $\mu$m, with a power of 20 $\mu$W in one-photon excitation, and 2.5 mW in two-photon excitation spectroscopy, except for Fig.~\ref{fig2}(c) where the power was decreased by a factor 10. An achromatic optical system couples the emitted signal to our detection system, composed of a f=500 mm Czerny-Turner monochromator, equipped with a 300 grooves/mm grating blazed at 250 nm, and with a back-illuminated CCD camera (Andor Newton 920), with a quantum efficiency of 50$\%$ at 210 nm, operated over integration times of 1 min.

\textbf{Acknowledgments}

We gratefully acknowledge C. L'Henoret, D. Rosales, and M. Moret for technical support, L. Tizei, O. Stephan, A. Zobelli, M. Kociak, L. Schue, J. Barjon, A. Loiseau, F. Ducastelle for fruitful discussions. This work was financially supported by the network GaNeX (ANR-11-LABX-0014). GaNeX belongs to the publicly funded \textit{Investissements d'Avenir} program managed by the French ANR agency. G.C. is a member of 'Institut Universitaire de France'.

$^\ast$e-mail: bernard.gil@umontpellier.fr

\end{document}